\documentclass[preprints,article,accept,pdftex,moreauthors]{Definitions/mdpi}

\firstpage{1}
\pubvolume{1}
\issuenum{1}
\articlenumber{0}
\pubyear{2026}
\copyrightyear{2026}
\datereceived{ }
\daterevised{ } 
\dateaccepted{ }
\datepublished{ }

\usepackage{aas_macros}

\Title{Disk-Like Matter Outflow from a Kerr-Type Wormhole: Possible Observational Manifestations and Magnetic Effects\\ {\small Accepted for publication in {\it Physics}}}

\Author{{Mikhail Piotrovich} *\orcidA{}, Stanislava Buliga \orcidB{} and Tinatin Natsvlishvili \orcidC{}}

\AuthorNames{Mikhail Piotrovich, Stanislava Buliga and Tinatin Natsvlishvili}

\address[1]{%
Central astronomical observatory at Pulkovo RAS, 196140 Saint-Petersburg, Russia; {aynim@yandex.ru (S.B.); tinatingao@mail.ru (T.N.)};}

\corres{Correspondence: mpiotrovich@mail.ru}

\abstract{We present a phenomenological study of a hypothetical compact object with a traversable 2 Kerr-type wormhole geometry connecting two asymptotically distinct regions. One mouth of the wormhole is assumed to reside in an active galactic nucleus (AGN) environment with ongoing accretion, while the second mouth is located in a comparatively quiescent galactic nucleus without accretion activity. In such a configuration, matter accreted near the AGN-side mouth may traverse the wormhole throat and emerge from the opposite mouth. Because the inflowing matter possesses angular momentum, the resulting outflow is expected to be predominantly equatorial, producing a disk-like structure fundamentally different from a conventional accretion disk around a black hole. Several dynamical regimes are considered depending on the outflow velocity relative to the local escape and orbital velocities. Possible observational manifestations, thermal properties, recombination signatures, and magnetic field effects are discussed. Special attention is given to the possibility of jet formation in the absence of a conventional accretion disk. We explicitly distinguish the weak-field, large-radius Newtonian classification used for the outer flow from the strong-field region near the throat, where the dynamics should be treated in a general-relativistic and magnetohydrodynamic framework. We further discuss the specific energy and angular momentum inherited from the inner edge of the accretion flow, the possible modification of these quantities by magnetic stresses and energy-extraction processes, and the effects of an intrinsic dipolar magnetic field of the wormhole.}


\begin{document}

\section{Introduction}

\subsection{Wormholes and Quantum Field Theory}

Traversable wormholes, originally formulated in the seminal work of Michael S. Morris and Kip S. Thorne \citep{morris88}, require the presence of so-called exotic matter capable of violating the null energy condition in order to stabilize the wormhole throat against gravitational collapse. A natural connection between wormhole physics and quantum field theory arises through the Casimir effect, in which quantum vacuum fluctuations between conducting boundaries generate a negative energy density. Since the Casimir effect provides an experimentally verified realization of negative vacuum energy, it is frequently regarded as one of the few physically plausible mechanisms that could, at least in principle, support traversable wormhole geometries. Recent theoretical studies have explored ''Casimir wormholes'', where semiclassical gravitational equations are sourced by Casimir-type vacuum energy distributions, suggesting that quantum vacuum effects may contribute to the stabilization of nontrivial spacetime topologies, although severe constraints on traversability and macroscopic stability remain unresolved \citep{garattini19,oliveira23}.

\subsection{Kerr-Type Traversable Wormholes}

Rotating traversable wormholes, whose exterior geometry resembles the Kerr metric, naturally arise in several extensions of general relativity and quantum gravity. In contrast to static Morris--Thorne wormholes, rotating configurations may inherit angular momentum either from the collapse of rotating compact objects, from rapidly rotating accretion flows, or from the dynamics of the exotic matter supporting the throat. One of the earliest approaches was proposed by Teo, who generalized the Morris--Thorne metric to include rotation while preserving traversability \citep{teo98}. In this framework, the wormhole angular momentum is introduced phenomenologically, leading to frame-dragging effects analogous to those around Kerr black holes.

Another possible formation mechanism involves gravitational collapse in modified theories of gravity. In scalar--tensor theories, Einstein--Gauss--Bonnet gravity, Einstein--Cartan gravity, and other generalized gravitational models, nontrivial scalar or torsion fields may prevent the formation of an event horizon while simultaneously allowing the remnant to retain a substantial fraction of the progenitor's angular momentum \citep{kanti11,bronnikov12}. Such rotating wormholes may therefore represent the endpoint of stellar or supermassive object evolution under conditions where exotic gravitational degrees of freedom become dynamically important.

Quantum gravity scenarios provide an additional class of formation mechanisms. In semiclassical gravity, quantum vacuum polarization and negative-energy fluctuations may stabilize microscopic wormholes predicted by spacetime foam models \citep{hochberg97,visser95}. If these microscopic structures are subjected to angular momentum transfer through interactions with surrounding matter or through mergers, they could evolve into rotating wormholes. Similar ideas have also been explored within loop quantum gravity and other approaches where quantum corrections replace classical singularities by nonsingular bridge-like geometries.

A further possibility is that rotating wormholes originate from the accretion-driven spin-up of initially non-rotating configurations. If a traversable wormhole is embedded in a binary system or at the center of a galaxy, prolonged accretion from a disk can efficiently transport angular momentum to the throat. Numerical studies indicate that such processes may produce rapidly rotating wormholes with observational signatures resembling those of Kerr black holes, while preserving distinctive features in their shadows, quasi-periodic oscillations, and polarization properties\citep{bambi13,shaikh18}. More specifically, recent accretion studies show that the near-throat flow can be intrinsically relativistic and magnetized: general-relativistic magnetohydrodynamic (GRMHD) simulations of accretion through a traversable wormhole find that inflowing plasma can accumulate in a hot, rotating cloud at the throat and drive an outflow from the opposite side \citep{combi24}. This result provides an important relativistic counterpart to the phenomenological picture considered below.

\subsection{Observational Signatures of Wormholes}

A considerable body of research has been devoted to the investigation of potential observational signatures of wormholes and to the identification of astrophysical phenomena that could reveal their existence \citep{harko08,harko09,tsukamoto12,yoo13,bambi13,bambi13b,dokuchaev14,li14,zhou16,dai19,paul20,tripathi20,bambi21,urtubey24,chen25}. In particular, numerous studies have focused on determining those observational characteristics that may enable a clear distinction between wormholes and black holes, since both classes of compact objects can exhibit similar gravitational effects and, consequently, may appear nearly indistinguishable in many astrophysical environments. The search for such distinguishing features has become an important topic in relativistic astrophysics, motivated by the possibility of testing alternative gravitational configurations and probing the fundamental structure of spacetime.

Despite these extensive efforts, the majority of previous investigations have reached a rather cautious conclusion. Although several potentially unique observational effects have been proposed, most of them are expected to be extremely subtle and therefore difficult to detect in practice. In many cases, the predicted signatures either lie below the sensitivity thresholds of current observational facilities or can be masked by astrophysical uncertainties and environmental effects. As a result, even when theoretical differences between wormholes and black holes are present, translating these differences into robust and unambiguous observational evidence remains a significant challenge. Consequently, the direct identification of wormholes through the majority of previously suggested methods appears to be beyond the capabilities of contemporary astronomical instrumentation or requires observational precision that is not yet achievable.

In the present work (as well as in our previous works with the participation of the late Serguei Krasnikov \citep{piotrovich20a,piotrovich20b,piotrovich24a}) we use a different, more phenomenological approach. Rather than concentrating on weak or highly model-dependent effects, we focus on astrophysical objects and physical mechanisms that are already well established and extensively studied within modern astrophysics. By considering processes whose properties, dynamics, and observational manifestations are well understood, we aim to maximize the likelihood of obtaining detectable and reliable signatures. This approach allows us to investigate scenarios in which the presence of a wormhole may produce pronounced observational consequences, potentially leading to strong and measurable effects that could be identified with existing or near-future observational facilities. Our objective is therefore to explore whether the combination of familiar astrophysical phenomena with wormhole geometries can generate observational manifestations that are significantly more prominent than those proposed in previous studies, thereby improving the prospects for an empirical distinction between wormholes and black holes.

\subsection{The Problem Considered in this Work}

Compact astrophysical objects associated with supermassive black holes are commonly interpreted within the framework of general relativity as Kerr black holes surrounded by accretion disks \citep{shakura73,novikov73}. However, various alternative compact objects have also been proposed, including traversable wormholes \citep{morris88,visser95,teo98}.

If traversable wormholes exist in nature, they could in principle connect distant astrophysical environments with dramatically different physical conditions. For example, there is a hypothesis that the central objects in the nuclei of galaxies may not be supermassive black holes, but wormholes \citep{kardashev07}. One may consider a configuration in which one mouth of a rotating wormhole is located in an active galactic nucleus (AGN), while the opposite mouth resides in a quiet galactic nucleus with negligible accretion activity.

In this scenario, matter accreted on the AGN side may pass through the wormhole throat and emerge from the second mouth. Since the accretion flow in AGNs is typically disk-like and possesses significant angular momentum, the emerging matter is also expected to form a predominantly equatorial outflow. Such a system may produce observational phenomena qualitatively different from ordinary black hole accretion disks. The purpose of this work is to investigate several possible regimes of disk-like outflows and discuss their potential observational signatures.

At first glance, the mechanisms we propose in this paper may seem fairly simple and obvious. However, as far as we can tell, they are all new and original. We were unable to find any references in the literature to similar ideas in the context of searching for observational manifestations of wormholes. We now qualify this statement in view of recent work: GRMHD simulations have independently demonstrated a throat-powered outflow in a traversable wormhole, although the present paper considers a different phenomenological configuration and focuses on the subsequent disk-like regimes and their observational consequences \citep{combi24}. The primary goal of this paper is therefore more narrowly to formulate a phenomenological framework and identify quantities that can be tested in future metric-specific calculations, rather than to claim that the proposed signatures are by themselves unique to wormholes.

The main difficulty in searching for wormholes is that we don't know (unlike black holes) what specific metric a wormhole should have. Therefore, we believe it's best to consider mechanisms that are minimally dependent on a specific metric. In this paper, we basically use only three wormhole characteristics: the mass, the presence of angular momentum and the absence of an event horizon. For the strong-field part of the problem, however, these three quantities are not sufficient to specify the actual plasma dynamics. In particular, the inner-edge specific energy and angular momentum depend on the stationary and axisymmetric metric, while the electromagnetic four-potential and magnetic stresses must be included when the flow is sufficiently ionized. We therefore regard the Newtonian treatment below as an outer-flow classification and explicitly identify the relativistic throat region as requiring a metric-specific GRMHD treatment. We emphasize that the absence of an event horizon can be precisely the most obvious observational manifestation of a wormhole.

\section{Physical Configuration}

\begin{figure}[ht!]
\centering
\includegraphics[width=\textwidth]{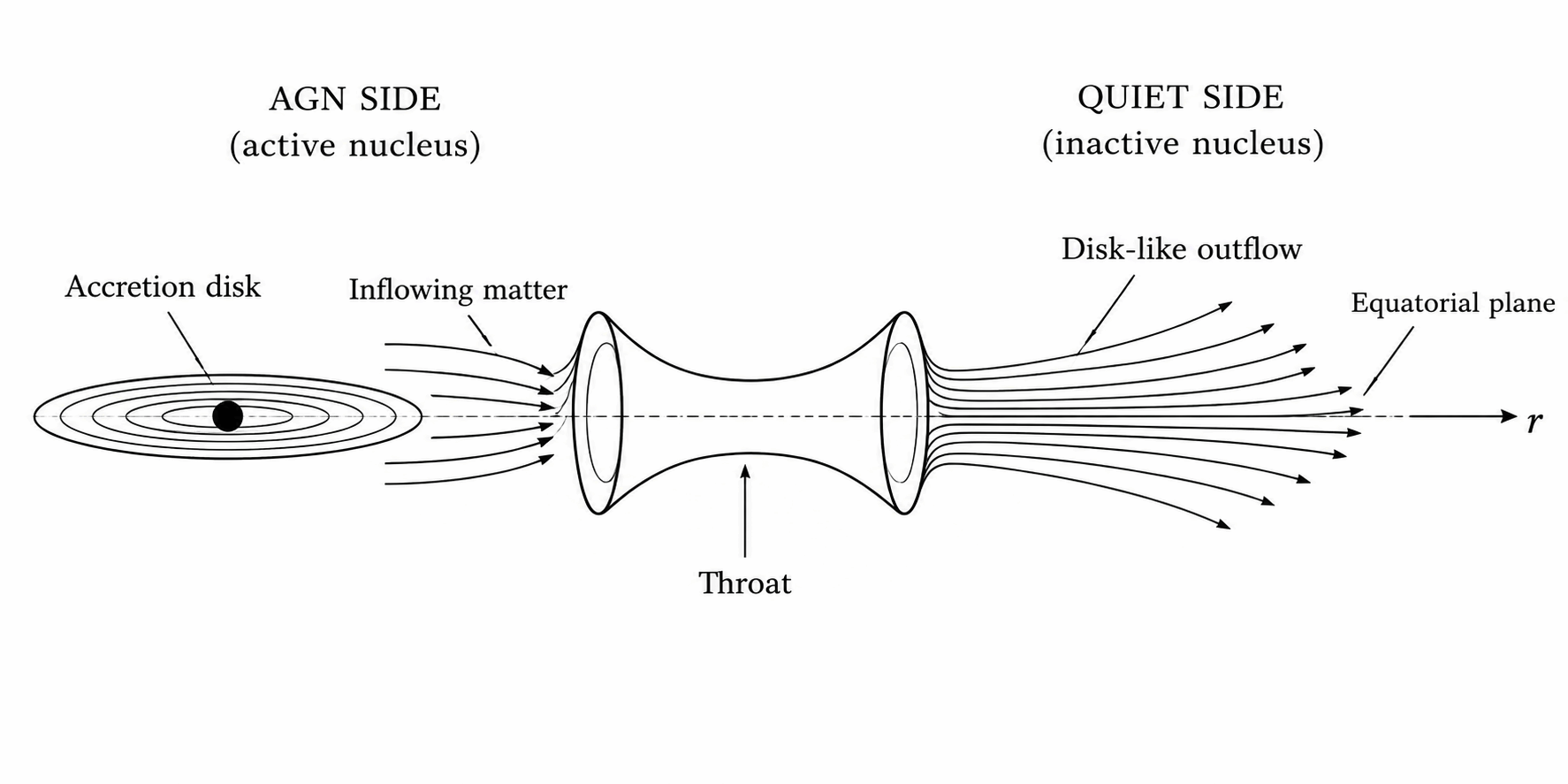}
\caption{\label{fig:configuration} Schematic configuration of the proposed system. The figure is not an embedding diagram of the wormhole equatorial plane; the two mouths and the throat are shown schematically, while the AGN-side accretion disk and the predominantly equatorial outflow from the quiet-side mouth are represented separately.}
\end{figure}

We consider a rotating compact object modeled phenomenologically as a Kerr-like traversable wormhole connecting two distinct astrophysical environments. In the scenario investigated in this work, one mouth of the wormhole is assumed to be located within the central region of an active galactic nucleus (AGN), where a dense accretion disk continuously supplies matter, angular momentum, and magnetic flux to the surrounding spacetime. The second mouth is placed in a comparatively quiescent galactic nucleus characterized by a significantly lower level of nuclear activity and a reduced gas supply (see Fig.~\ref{fig:configuration}). Such a configuration provides a natural mechanism for the transport of matter and energy from an active environment to an otherwise inactive galactic center through the wormhole throat.

The accreting material entering the wormhole from the AGN side is expected to possess substantial angular momentum inherited from the rotational motion of the accretion disk. In addition, the plasma is likely to carry large-scale magnetic fields that are frozen into the conducting flow. As a consequence, the matter emerging from the second mouth cannot be regarded as a spherically symmetric outflow. Instead, the rotational properties of the wormhole geometry and the angular momentum of the transported plasma naturally lead to a highly anisotropic matter distribution. The outflow is expected to be concentrated predominantly in the vicinity of the equatorial plane of the rotating wormhole mouth, where centrifugal effects are most significant. Furthermore, the transported magnetic field may influence the subsequent evolution of the emerging gas through magnetic pressure, magnetic tension, and magnetohydrodynamic instabilities.

The physical state of the matter immediately before and after the throat should, however, be understood as a strong-field problem. In a stationary, axisymmetric geometry the accreting plasma is characterized by conserved specific energy $E$ and specific angular momentum $L$ along geodesic motion, while the inner edge of a thin disk is associated with the marginally stable circular orbit (when such an orbit exists). Relativistic calculations of rotating wormhole disks explicitly determine the marginally stable orbit, specific energy, specific angular momentum, and accretion efficiency from the metric \citep{rahaman21}. Thus the quantities carried through the throat should not in general be identified with values determined by the outer, nearly Keplerian part of the disk. Instead, in the simplest geodesic limit we may write
\begin{equation}
\boldsymbol{E_{\rm in}\simeq E_{\rm ms}, \qquad L_{\rm in}\simeq L_{\rm ms},}
\end{equation}
where the subscript ``ms'' denotes the marginally stable circular orbit. This relation is a limiting approximation: magnetic stresses, pressure forces, non-geodesic motion, and the detailed throat geometry can shift the effective inner-edge values.

For an ionized flow, the appropriate relativistic equations are the ideal-GRMHD conservation laws, schematically
\begin{align}
\boldsymbol{\nabla_\mu(\rho u^\mu)}&=0,\\
\boldsymbol{\nabla_\mu T^{\mu\nu}}&=0,\\
\boldsymbol{\nabla_\mu {}^*F^{\mu\nu}}&=0,
\end{align}
with $T^{\mu\nu}$ containing both matter and electromagnetic contributions. These equations make explicit why the magnetic field and frame dragging cannot be neglected close to the throat. We do not attempt a metric-specific solution of these equations here; instead, we use them to delimit the regime of validity of the phenomenological treatment. Recent GRMHD simulations provide a concrete example in which magnetized inflowing plasma forms a hot rotating throat cloud and launches a mildly relativistic thermal wind \citep{combi24}.

For the simplest Newtonian bookkeeping on the two sides, the energy budget can be written as\begin{equation}\frac{1}{2}v_{\rm out}^{2}=\frac{1}{2}v_{\infty}^{2}+\frac{GM_1}{r_1}-\frac{GM_2}{r_2}+W_{\rm throat},\end{equation}where $M_1$ and $M_2$ are the effective masses associated with the active and quiet mouths, respectively, $r_1$ and $r_2$ are the corresponding reference radii, and $W_{\rm throat}$ represents net specific work performed by the throat region, including possible electromagnetic or rotational energy transfer. For equal masses and approximately symmetric radii, with negligible throat work, one obtains $v_{\rm out}\simeq v_{\infty}$; hence a Regime-I outflow is not obtained automatically from symmetric free fall. It requires either an asymmetric gravitational potential, positive throat work, or another energy-extraction process.

The dynamical evolution of the matter after its emergence from the wormhole depends on the interplay between the gravitational field of the compact object and the kinematic properties of the outflow. In particular, the characteristic velocities $v_{\rm out}$, $v_{\rm circ}$, and $v_{\rm esc}$ remain useful reference scales for the outer flow. They are no longer used as a three-way dynamical classification: the sign of the specific energy determines whether the orbit is bound or unbound, while the specific angular momentum and the throat radius determine the subsequent return and circularization behavior.

The local escape velocity in the Newtonian limit is
\begin{equation}
v_{\rm esc}=\sqrt{\frac{2GM}{r}},
\end{equation}
where $G$ is the gravitational constant, $M$ denotes the effective gravitational mass associated with the quiet-side mouth, and $r$ is the radial distance from its center. The local circular velocity is
\begin{equation}
v_{\rm circ}=\sqrt{\frac{GM}{r}}.
\end{equation}
These expressions are retained only as outer-flow reference scales; in the strong-field region the corresponding quantities must be obtained from the relativistic metric and the plasma equations.

These dynamical outcomes are expected to have important observational consequences. The accumulation of matter around the inactive-side mouth may lead to enhanced gas densities, the formation of long-lived accretion structures, and the development of secondary episodes of nuclear activity. At the same time, the transport of magnetic flux through the wormhole could facilitate the generation of magnetized outflows and potentially contribute to jet formation. Therefore, the study of the velocity distribution of the emerging matter represents a key step toward understanding the observable manifestations of traversable wormholes embedded in realistic astrophysical environments.

\section{Dynamics and Physical Parameters of Matter Near a Wormhole}

The physical evolution of matter emerging from the inactive-side mouth of the wormhole is determined primarily by the interplay between gravitational forces, angular momentum, magnetic stresses, and radiative processes. Since the plasma originates from an accretion flow in an active galactic nucleus, it is expected to carry substantial kinetic energy, angular momentum, and magnetic flux. The subsequent evolution of this material depends critically on its velocity distribution at the moment of emergence from the wormhole throat.

The three outcomes considered below are now classified in terms of the specific orbital energy and angular momentum rather than by $v_{\rm out}$ alone. This removes the artificial boundary at $v_{\rm circ}$ and makes explicit that orbital type is determined by two quantities. In the weak-field Newtonian limit, at an emergence radius $b_0$ on the quiet side, we define
\begin{equation}
\epsilon=\frac{v_r^2+v_\phi^2}{2}-\frac{GM_2}{b_0}, \qquad \ell=b_0v_\phi.
\end{equation}
Here $E$ and $L$ in the relativistic notation denote the conserved specific energy and specific angular momentum, while $\epsilon$ and $\ell$ are their Newtonian outer-flow counterparts. Bound motion has $\epsilon<0$, whereas $\epsilon>0$ corresponds to unbound hyperbolic motion. For $\epsilon<0$, the Newtonian eccentricity is
\begin{equation}
e=\sqrt{1+\frac{2\epsilon\ell^2}{(GM_2)^2}}.
\end{equation}
The role of the throat radius is also explicit: a bound stream launched outward from $b_0$ can return to the throat unless pressure, shocks, magnetic stresses, or other dissipation circularize it before the first return. Thus the distinction between the two bound outcomes is dynamical, not a consequence of the inequality $v_{\rm out}\lessgtr v_{\rm circ}$.

\begin{figure}[ht!]
\centering
\includegraphics[width=0.8\textwidth]{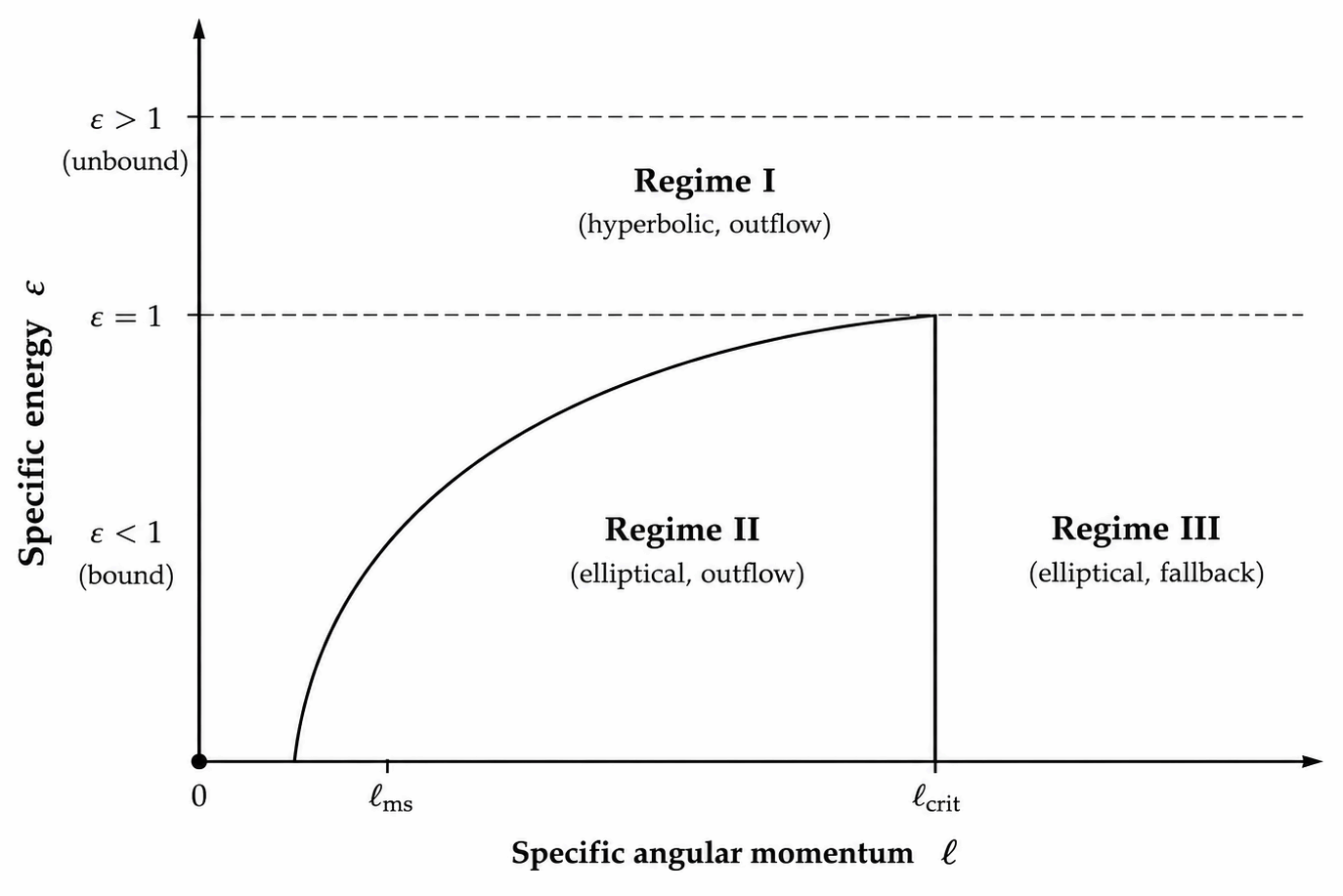}
\caption{Schematic classification of the outer flow in the $(\epsilon,\ell)$ plane; the figure is illustrative rather than a metric-specific calculation.}
\label{fig:regimes}
\end{figure}

\subsection{Regime I: Unbound Hyperbolic Outflow}

The first regime corresponds to the situation in which the kinetic energy of the emerging plasma is sufficiently large to overcome the gravitational attraction of the wormhole mouth. This may occur when the effective gravitational influence of the AGN-side mouth is comparable to or exceeds that of the quiet-side mouth, allowing the transported matter to emerge with a large velocity.

The defining condition is
\begin{equation}
\epsilon>0,
\end{equation}
or, equivalently in the purely radial Newtonian limit, $v_{\rm out}>v_{\rm esc}$. Under these circumstances, the plasma follows unbound trajectories and escapes from the gravitational potential of the compact object.

Although the emerging material originates predominantly from the equatorial region of the rotating wormhole mouth, its subsequent motion differs fundamentally from that of matter in a conventional accretion disk. Owing to the anisotropic geometry of the outflow, the resulting structure may appear observationally similar to a flattened disk-like configuration. However, the underlying dynamics are radically different because the plasma is moving outward rather than inward.

In a standard accretion disk surrounding a black hole or other compact object, angular momentum support leads to predominantly azimuthal motion. Consequently,
\begin{equation}
v_\phi\gg v_r,
\end{equation}
where $v_\phi$ and $v_r$ denote the azimuthal and radial velocity components, respectively.

\begin{figure}[ht!]
\centering
\includegraphics[width=0.6\textwidth]{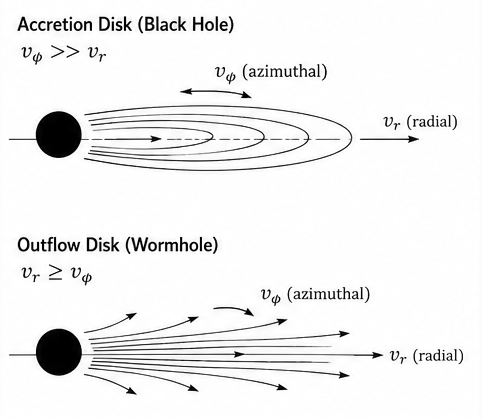}
\caption{Schematic radial and azimuthal velocity components.}
\label{fig:velocities}
\end{figure}

The situation is markedly different for matter emerging from a wormhole in the hyperbolic-outflow regime. Since the plasma is injected outward with significant kinetic energy, the radial component of the velocity can become comparable to or even exceed the rotational component:
\begin{equation}
v_r\gtrsim v_\phi.
\end{equation}
Such a velocity structure implies a fundamentally different mechanism of energy transport, density evolution, and radiative cooling. The reduced importance of viscous angular-momentum transport may substantially alter the thermal structure of the flow compared with ordinary accretion disks.

\begin{figure}[ht!]
\centering
\includegraphics[width=0.8\textwidth]{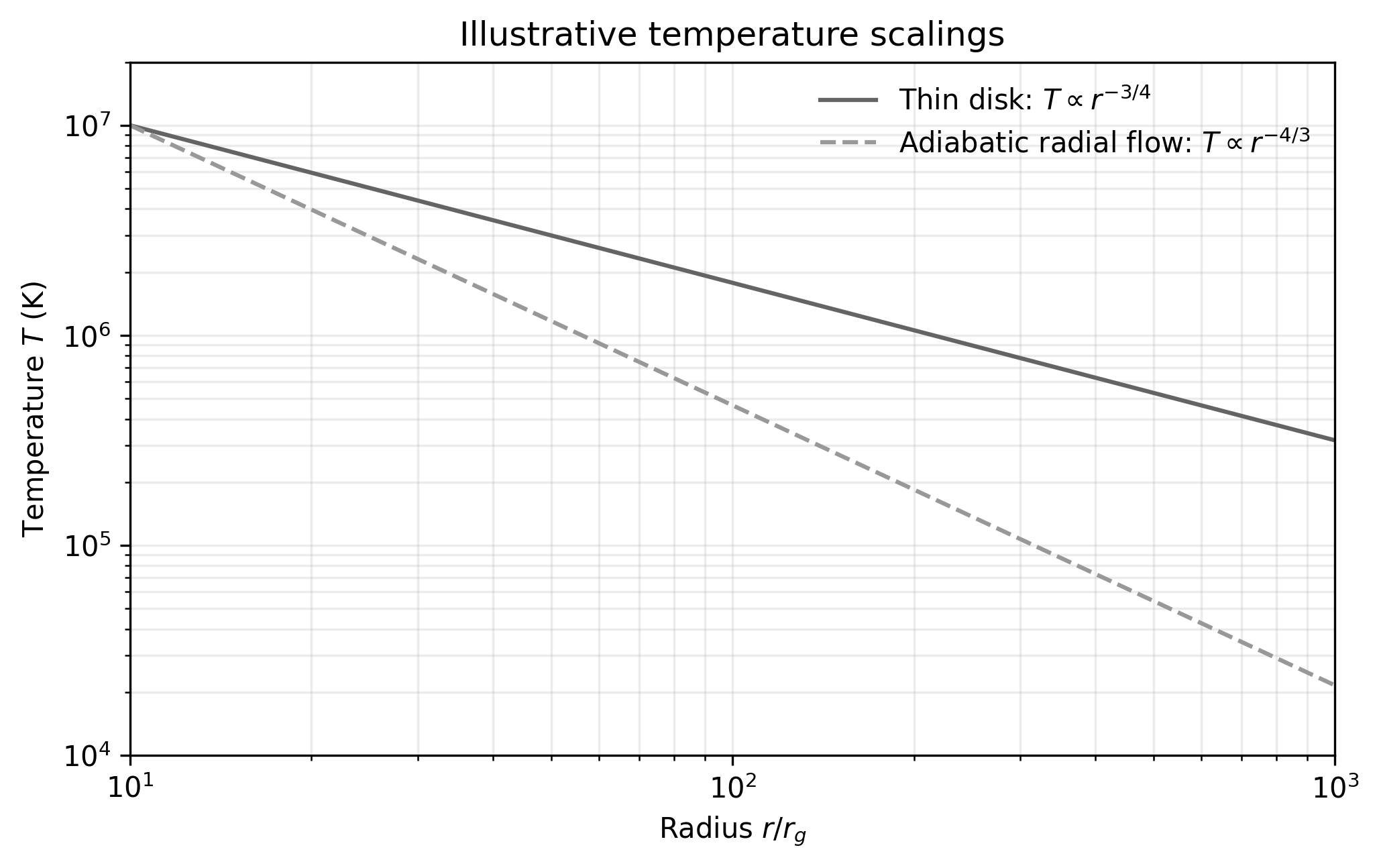}
\caption{Illustrative temperature scalings with numerical axes; the adiabatic-flow curve is shown to emphasize that a non-standard slope is not obtained without specifying the thermodynamic regime.}
\label{fig:temperature}
\end{figure}

For comparison, the effective temperature profile of a standard geometrically thin and optically thick accretion disk is approximately given by \citep{shakura73}
\begin{equation}
T(r)\propto r^{-3/4}.
\end{equation}
The full thin-disk expression also contains the usual inner-boundary factor $[1-(r_{\rm in}/r)^{1/2}]^{1/4}$ \citep{shakura73,novikov73}. This relation results from viscous dissipation within a nearly Keplerian Shakura--Sunyaev disk. In contrast, an outflow-dominated configuration is governed by expansion, adiabatic cooling, shocks, and radiative losses rather than inward accretion. For example, if the outer flow is approximately radial with nearly constant $v_r$, continuity gives $\rho\propto r^{-2}$ and adiabatic expansion with $\gamma=5/3$ gives $T\propto\rho^{\gamma-1}\propto r^{-4/3}$. Thus the original assumption $T\propto r^{-p}$ with $p<3/4$ is not generic and should not be used as a prediction without an explicit heating or photoionization model.

As the plasma expands to large distances from the wormhole mouth, its density gradually decreases. The corresponding reduction in collision rates eventually permits radiative recombination processes to become efficient. Under suitable physical conditions, this may lead to the appearance of a pronounced Radiative Recombination Continuum (RRC) feature in the observed spectrum \citep{liedahl90,kallman01}. The strength of this component depends on the emission measure and ionization state and is therefore estimated below rather than assumed to be large.

Furthermore, shocks generated by velocity gradients within the outflow may produce localized heating and particle acceleration. Such processes may enhance non-thermal radiation, generating additional emission in the ultraviolet, X-ray, and possibly even gamma-ray bands. Therefore, the spectral appearance of this regime could differ dramatically from that of ordinary supermassive black hole systems. These spectral differences are not unique to wormholes, however, and must be compared with non-standard black hole accretion flows.

\subsection{Regime II: Bound Elliptical Outflow and Circularization}

A second outcome corresponds to bound motion with $\epsilon<0$ and substantial angular momentum. The material initially follows eccentric trajectories, but shocks, pressure forces, magnetic stresses, and radiative losses may dissipate orbital energy before the first return to the throat and produce a rotating, gradually circularizing structure. The appropriate condition is therefore not $v_{\rm circ}\le v_{\rm out}<v_{\rm esc}$, but rather
\begin{equation}
\epsilon<0, \qquad t_{\rm diss}<t_{\rm ret},
\end{equation}
where $t_{\rm diss}$ is the characteristic dissipation/circularization time and $t_{\rm ret}$ is the first-return time to the throat.

In a standard Shakura--Sunyaev accretion disk, the radial velocity of the accreting matter is significantly (by orders of magnitude) less than the azimuthal velocity. In our case, however, the radial velocity can initially be comparable to the azimuthal velocity. The newly formed structure is expected to contain a broad distribution of eccentric trajectories. The eccentricity is given by the expression above in the Newtonian limit; in a relativistic treatment it must instead be obtained from the geodesic or MHD integrals of motion.

Initially, many of the resulting trajectories may possess large eccentricities. Consequently, different orbital streams can intersect one another, producing shocks, turbulent mixing, and dissipation of kinetic energy. These interactions naturally lead to the conversion of ordered orbital motion into thermal energy. This type of eccentric, self-intersecting circularization is also encountered in tidal-disruption-event debris, illustrating that the proposed phenomenology is not unique to wormholes \citep{hayasaki13}.

The resulting disk-like structure is therefore expected to be highly dynamic. Large-scale turbulence, density inhomogeneities, spiral features, and transient shock fronts may all be present during the early stages of evolution. The disk may exhibit significant variability across a wide range of timescales, reflecting the complex interplay between gravitational dynamics and hydrodynamical dissipation.

As the system evolves, frictional interactions and turbulent viscosity gradually reduce the orbital eccentricities of individual particles. Energy is radiated away through various cooling processes, including thermal emission, line radiation, and inverse Compton scattering. The progressive loss of orbital energy causes the disk to contract toward smaller radii.

Several important evolutionary trends are therefore expected: the average orbital eccentricity decreases with time, the characteristic disk radius gradually shrinks, the velocity distribution approaches a more Keplerian form, shock activity becomes less pronounced, the spectral energy distribution evolves continuously, and the thermal luminosity changes as the disk circularizes.

Eventually, if the matter supply from the wormhole ceases, the remnant disk may approach a configuration resembling a conventional accretion disk. Nevertheless, traces of its unusual formation history may persist in the form of residual eccentricity, altered chemical composition, non-standard density profiles, or distinctive spectral features. The transition from a highly eccentric turbulent disk to a quasi-Keplerian configuration may itself constitute an observable phenomenon, particularly if it occurs on timescales accessible to astronomical monitoring campaigns.

\subsection{Regime III: Bound Fallback and Plasma Shell}

The third outcome corresponds to bound material with insufficient dissipation or angular-momentum support to circularize before returning to the throat. It is characterized by
\begin{equation}
\epsilon<0, \qquad t_{\rm diss}\gtrsim t_{\rm ret},
\end{equation}
rather than by the condition $v_{\rm out}<v_{\rm circ}$. A small specific angular momentum makes the return especially direct, but the relevant boundary depends on $b_0$ and on the relativistic effective potential.

The resulting trajectories resemble ballistic motions rather than true orbital paths. Since the material remains concentrated near the equatorial plane, repeated episodes of ejection and fallback may lead to the formation of a dense plasma shell surrounding the compact object. For a Newtonian estimate the free-fall/dynamical timescale is
\begin{equation}
t_{\rm ff}\sim\sqrt{\frac{r^3}{GM}},
\end{equation}
which determines the typical duration over which matter returns to the vicinity of the wormhole mouth after its initial emergence.

Because the plasma remains confined to a relatively small spatial volume, its density can become substantially larger than in the previous two regimes. Gravitational compression and repeated collisions between infalling and newly emerging material are expected to generate strong heating. As a result, the plasma temperature may reach values characteristic of hot X-ray emitting environments.

The physical state of the shell is likely to be highly complex. Hydrodynamic instabilities, magnetic reconnection events, and turbulent motions may continuously redistribute energy throughout the system. Depending on the magnetic field strength transported through the wormhole, magnetically driven flares analogous to those observed in accreting compact objects may also occur.

The dominant radiative mechanisms in this regime are expected to differ significantly from those operating in conventional accretion disks. Because the plasma does not form a large, optically thick disk, the familiar multi-temperature black-body spectrum may be absent. Instead, the emitted radiation may be dominated by processes such as thermal bremsstrahlung, inverse Compton scattering, recombination radiation, and emission lines from highly ionized species \citep{rybicki79}.

Consequently, the observed spectrum may exhibit strong X-ray emission, prominent ionized iron lines, and signatures of hot optically thin plasma. Variability associated with fallback cycles and magnetic activity could provide additional observational diagnostics. Such diagnostics are likewise degenerate with radiatively inefficient accretion flows around black holes, which can produce hot, optically thin, non-Keplerian structures and outflows \citep{hada24,balakrishnan24}.

Overall, the suborbital-shell regime represents the most compact and thermally extreme configuration among the three scenarios considered here. It should be regarded as a possible fallback outcome rather than as a separate orbital class fixed by $v_{\rm circ}$.

\subsection{On the Applicability of the Newtonian Approximation}

The Newtonian approximation is used here only to classify the large-radius outer flow after the relativistic injection conditions have been established. It is not intended to describe the throat itself. There is no sharp radius at which relativistic effects suddenly become negligible; in particular, relativistic orbital dynamics can influence line profiles and other observables outside the innermost few gravitational radii. The strong-field region and the inner-edge boundary conditions must be treated with a metric-specific relativistic calculation, while the asymptotic flow can be described by the Newtonian expressions used above.

\subsection{Fiducial Numerical Scales}

To provide an order-of-magnitude scale, consider a fiducial wormhole mouth with $M_2=10^8M_\odot$, throat/emergence radius $b_0=20r_g$, where $r_g=GM_2/c^2$, and mass transfer rate $\dot M=0.1M_\odot\,\mathrm{yr}^{-1}$. At $20r_g$, the Newtonian reference velocities are $v_{\rm circ}\simeq0.224c$ and $v_{\rm esc}\simeq0.316c$. Illustrative values $v_{\rm out}=0.35c$, $0.25c$, and $0.10c$ correspond respectively to $\epsilon/c^2\simeq0.011$, $-0.019$, and $-0.045$ for a purely tangential launch and therefore illustrate unbound, bound, and strongly bound cases. They are not predictions for a particular wormhole metric.

The corresponding kinetic powers, $L_{\rm kin}=\dot M v_{\rm out}^2/2$, are approximately $3.5\times10^{44}$, $1.8\times10^{44}$, and $2.8\times10^{43}\,\mathrm{erg\,s^{-1}}$, respectively. If an illustrative fraction $\eta_{\rm rad}=0.01$--$0.1$ of this kinetic power is radiated, the associated radiative scale is $\sim10^{41}$--$10^{43}\,\mathrm{erg\,s^{-1}}$; a self-consistent value of $\eta_{\rm rad}$ requires radiative transfer. The Eddington luminosity of the fiducial $10^8M_\odot$ object is $L_{\rm Edd}\simeq1.3\times10^{46}\,\mathrm{erg\,s^{-1}}$.

For a quasi-steady radial component $v_r=0.1c$, continuity gives
\begin{equation}
\rho(r)=\frac{\dot M}{4\pi r^2v_r}, \qquad \rho(20r_g)\simeq1.9\times10^{-15}\,\mathrm{g\,cm^{-3}},
\end{equation}
corresponding to $n\sim10^9\,\mathrm{cm^{-3}}$ for ionized hydrogen. The associated flow time at $20r_g$ is about $1.1$ days for $v_r=0.1c$. As a simple emission-measure estimate, an equatorial emitting volume $V\sim0.1(4\pi r^3)$ at this radius gives $V\sim3\times10^{43}\,\mathrm{cm^3}$. Taking a representative recombination coefficient $\alpha_{\rm eff}\sim10^{-13}\,\mathrm{cm^3\,s^{-1}}$ and photon energy $h\nu\sim10^{-11}$ erg yields $L_{\rm rec}\sim\alpha_{\rm eff}n^2Vh\nu\sim3\times10^{37}\,\mathrm{erg\,s^{-1}}$, before ion-fraction and radiative-transfer corrections. This estimate demonstrates explicitly that the RRC strength is controlled by the density and emitting volume rather than being a free qualitative signature.

Finally, if the active-side field is $B_1=10^3$ G at $r_1=10r_g$ and the field is approximately radial, flux conservation gives $B_2\simeq B_1(r_1/r_2)^2$. Thus at $r_2=20r_g$ one obtains $B_2\sim250$ G, while at $100r_g$ one obtains $B_2\sim10$ G. These values are illustrative scaling estimates, not metric-specific predictions.

\section{Magnetic Fields}

We use Gaussian cgs units in this section, as indicated by the appearance of $8\pi$ in the magnetic-energy density. Magnetic fields are expected to play a fundamental role in the dynamics, energetics, and observational appearance of matter emerging from a traversable wormhole. Since the plasma transported through the wormhole originates in the vicinity of an active galactic nucleus, where strong magnetic fields are naturally generated and amplified by accretion processes, it is reasonable to expect that a significant amount of magnetic flux may accompany the transferred matter. The magnetic field may influence the evolution of the outflow on both local and global scales, affecting angular momentum transport, plasma confinement, shock formation, particle acceleration, and radiative emission.

The magnetic environment near the inactive-side mouth of the wormhole may therefore differ substantially from that expected around an isolated compact object located in a quiescent galactic nucleus. In particular, the wormhole may serve not only as a channel for matter transport but also as a conduit through which magnetic flux generated in one astrophysical environment is deposited into another. Such a process could produce observable magnetic activity in regions where no conventional mechanism exists to sustain strong magnetic fields.

\begin{figure}[ht!]
\centering\includegraphics[width=\textwidth]{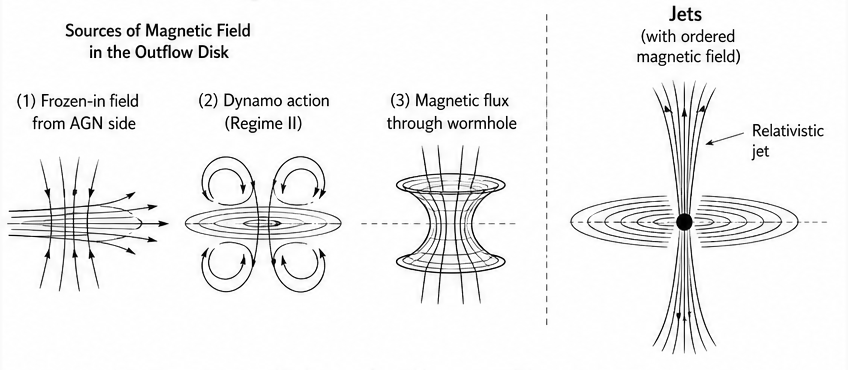}
\caption{Schematic magnetic-field transport, amplification, and jet/outflow possibilities.}
\label{fig:magnetic}
\end{figure}

Several distinct mechanisms may contribute to the generation, amplification, and transport of magnetic fields in disk-like structures associated with wormhole outflows. Although these mechanisms are discussed separately below, they are not mutually exclusive and may operate simultaneously in realistic astrophysical systems (see Fig.~\ref{fig:magnetic}).

\subsection{Frozen-In Magnetic Flux}

The most natural source of magnetic fields in the emerging plasma is the magnetic flux already present in the accretion disk surrounding the active-side wormhole mouth. Observations and theoretical models of active galactic nuclei indicate that accretion disks are strongly magnetized environments in which magnetic fields participate directly in angular momentum transport and energy dissipation.

When ionized plasma enters the wormhole throat, the magnetic field embedded within the plasma may be transported together with the matter flow. Under the assumption of ideal magnetohydrodynamics (MHD), the electrical conductivity of the plasma is sufficiently high that magnetic field lines remain effectively attached to the moving fluid. The temporal evolution of the magnetic field is then governed by the induction equation
\begin{equation}
\frac{\partial\mathbf B}{\partial t}=\nabla\times(\mathbf v\times\mathbf B),
\end{equation}
where $\mathbf B$ denotes the magnetic field vector and $\mathbf v$ is the plasma velocity field. This equation expresses the well-known frozen-in flux condition, according to which magnetic field lines are advected together with the plasma. Consequently, the magnetic topology generated in the AGN environment may survive the passage through the wormhole and reappear near the second mouth.

The subsequent evolution of the transported field depends strongly on the geometry of the emerging flow. In the case of a radially expanding outflow, magnetic flux conservation implies a gradual decrease of the field strength with increasing distance from the compact object. For a predominantly radial magnetic configuration, one expects approximately
\begin{equation}
B_r\propto r^{-2},
\end{equation}
whereas a toroidal component generated by rotational motion may decrease more slowly,
\begin{equation}
B_\phi\propto r^{-1}.
\end{equation}
As a result, the large-scale magnetic structure of the outflow may evolve from a mixed poloidal-toroidal configuration near the wormhole mouth to a toroidally dominated field at larger distances.

The presence of frozen-in magnetic fields can have important dynamical consequences. Magnetic pressure contributes to the overall force balance of the plasma and may influence the collimation of the outflow. In addition, magnetic tension forces can alter the trajectories of charged particles and affect the development of hydrodynamic instabilities. Strong magnetic fields may also provide an additional source of support against gravitational compression, particularly in the dense inner regions of the flow.

Another important consequence is the possibility of magnetic reconnection. As the plasma expands and the magnetic field becomes distorted by turbulence and differential motion, oppositely directed field lines may come into contact and reconnect. Such reconnection events convert magnetic energy into thermal energy and accelerated particles, potentially producing transient flares observable in radio, optical, ultraviolet, or X-ray bands.

Furthermore, the transport of frozen-in magnetic flux through the wormhole may provide a mechanism for establishing a coherent large-scale magnetic field in an otherwise weakly magnetized galactic nucleus. Such a process could significantly alter the long-term evolution of the surrounding interstellar medium.

\subsection{Dynamo Processes}

The frozen-in and dynamo descriptions should not be interpreted as simultaneous requirements. In a strongly magnetized, low-$\beta$ flow the imposed field can suppress or modify MRI growth, whereas in a weakly magnetized flow with $\beta=8\pi p/B^2\gg1$, differential rotation can allow MRI-driven amplification. We therefore treat the two mechanisms as alternative limiting magnetic regimes. In addition to the direct transport of magnetic fields from the AGN side, magnetic energy may also be generated locally after the plasma emerges from the wormhole. This process is particularly relevant in the bound-disk regime discussed previously, where the outflow settles into a rotating and turbulent disk-like structure.

Theoretical studies of accretion disks have demonstrated that turbulence combined with differential rotation can drive efficient magnetic-field amplification through dynamo action. One of the most important mechanisms responsible for the onset of turbulence in magnetized disks is the magnetorotational instability (MRI), originally investigated by Balbus and Hawley \citep{balbus91}. The MRI efficiently converts rotational energy into turbulent motion and provides a natural mechanism for magnetic-field growth. The characteristic condition for MRI development is
\begin{equation}
\frac{d\Omega}{dr}<0,
\end{equation}
where $\Omega$ is the angular velocity of the disk material.

Because the bound-disk regime around the wormhole mouth is expected to exhibit substantial differential rotation, the conditions for MRI-driven turbulence may naturally arise. Once turbulence develops, magnetic field lines become stretched, twisted, and folded by the fluid motions. These processes convert kinetic energy into magnetic energy and can amplify initially weak seed fields by many orders of magnitude.

The resulting dynamo activity may generate both small-scale turbulent magnetic structures and large-scale coherent magnetic configurations. Depending on the physical properties of the disk, the amplified field may eventually approach equipartition with the turbulent kinetic energy,
\begin{equation}
\frac{B^2}{8\pi}\sim\frac{1}{2}\rho v_{\rm turb}^2,
\end{equation}
where $\rho$ is the plasma density and $v_{\rm turb}$ is the characteristic turbulent velocity.

Such strong magnetic fields may significantly influence the subsequent evolution of the disk. Magnetic stresses contribute to angular momentum transport, allowing matter to migrate inward while angular momentum is transferred outward. At the same time, magnetic buoyancy may cause field lines to rise above the disk surface, forming a magnetized corona analogous to those observed in conventional accretion systems. The presence of a corona may enhance high-energy radiation through inverse Compton scattering of lower-energy photons. Consequently, dynamo-generated magnetic fields may leave detectable signatures in the X-ray spectra of wormhole-associated disks.

An additional consequence of dynamo amplification is the possibility of cyclic magnetic activity. Similar to the magnetic cycles observed in stars and galaxies, large-scale field reversals could occur over long timescales, leading to quasi-periodic variations in luminosity, polarization, and jet activity.

\subsection{Direct Magnetic Field Penetration}

A third and particularly intriguing possibility is the direct transmission of magnetic flux through the wormhole throat itself. In this scenario, the wormhole acts not merely as a conduit for magnetized plasma but as a channel through which magnetic field lines extend continuously between the two mouths.

If such a configuration is physically realized, large-scale magnetic structures generated in the vicinity of the AGN-side accretion disk may emerge near the quiet-side mouth with relatively little modification (see Fig.~\ref{fig:magnetic}). This possibility is supported at the phenomenological level by the idea that wormhole-like electromagnetic channels can transport magnetic flux between distant regions, although the astrophysical gravitational case requires a self-consistent relativistic treatment.

This possibility has several remarkable implications. First, the magnetic energy available near the quiet-side mouth may greatly exceed that expected from local astrophysical processes. Strong ordered magnetic fields could therefore appear in galactic nuclei that otherwise show little evidence of recent accretion activity. Second, magnetic stresses may provide a mechanism for transferring energy and angular momentum between the two environments connected by the wormhole. In principle, variations occurring near the AGN-side mouth could influence the magnetic configuration near the opposite mouth, producing correlated variability between regions separated by enormous distances in ordinary spacetime.

Third, large-scale magnetic flux emerging from the wormhole could facilitate the formation of relativistic jets (see the right side of Fig.~\ref{fig:magnetic}). In many models of jet production, ordered magnetic fields are required to extract rotational energy and accelerate plasma to high velocities. In standard black hole systems, jets are usually associated with accretion disks and magnetically driven extraction of rotational energy \citep{blandford77,blandford82}. Note that in the black hole case the magnetic field is generated only by the accretion disk. In the case of wormholes with an accretion disk, a somewhat similar scenario may also theoretically occur for some types of wormholes with a high spin value \citep{urtubey24}. However, a jet near the wormhole with magnetic field might also exist even without a directly observable accretion disk.

Some jet formation mechanisms \citep{blandford77} require magnetic fields and surrounding plasma but do not necessarily require continuous local accretion onto the compact object. In this case, the energy for creating the jet will be extracted from the rotational energy of the wormhole. Consequently, a wormhole system could potentially exhibit relativistic jets emerging from an apparently inactive galactic nucleus. Such an observational signature would be difficult to reconcile with standard black hole models.

The observational signatures of such a process could be highly unusual. For example, one might observe powerful radio emission, polarized synchrotron radiation, or collimated outflows emerging from a galactic nucleus that exhibits little evidence for the accretion activity normally required to sustain such phenomena. The coexistence of strong magnetic activity with an apparently inactive environment could therefore constitute a potential observational indicator of wormhole-mediated magnetic flux transport.

More generally, direct magnetic field penetration through the wormhole would establish a unique physical connection between two distant astrophysical systems. The resulting magnetospheric structure could possess properties unlike those encountered in conventional compact objects and may provide some of the most distinctive observational signatures of traversable wormholes embedded in realistic galactic environments.

A further possibility, which is particularly relevant to the present model, is an intrinsic magnetic field associated with the wormhole itself. Slowly rotating magnetized wormholes with a non-vanishing magnetic moment have been studied explicitly, including the electromagnetic potentials and the motion of charged particles in their vicinity \citep{abdujabbarov09}. More recent work has shown that an internal dipole field can also be considered for black hole mimickers such as wormholes, with the dipole source located inside the compact object and with the strong-field particle dynamics determined by the coupled gravitational and electromagnetic fields \citep{stuchlik26}. We therefore consider a simple phenomenological dipole field as an additional component of the magnetic environment. In the weak-field, large-radius limit its magnitude can be represented schematically as
\begin{equation}
\boldsymbol{B_{\rm dip}(r,\theta)\sim\frac{\mu}{r^3}\left(2\cos\theta\,\hat{\mathbf r}+\sin\theta\,\hat{\boldsymbol\theta}\right),}
\end{equation}
where $\mu$ is an effective magnetic dipole moment. This expression is not intended to represent an exact solution for a rotating wormhole metric; it is a controlled phenomenological limit. Close to the throat, the actual field must be obtained by solving Maxwell's equations in the chosen wormhole geometry, and its interaction with plasma must be included in the GRMHD equations. The intrinsic field can modify the inner-edge angular momentum and energy through Lorentz forces, change the location of stable circular orbits for charged plasma, and alter the conditions for reconnection and jet formation. The existence of such a field therefore provides an additional reason why the specific angular momentum transported through the throat should not be assumed to be purely geodesic.

In addition, rotating wormhole magnetospheres have recently been treated directly in general relativity. Ertola Urtubey, Perez, and Romero found that a Kerr-type rotating wormhole can produce an outgoing electromagnetic Poynting flux when accreting magnetized matter, with a mechanism analogous to Blandford--Znajek for rapidly rotating configurations \citep{urtubey24}. A subsequent calculation of the magnetic field geometry in a rotating wormhole solved the force-free stream equation and found that deviations from the Kerr magnetic configuration appear at second order in the spin and increase with the deviation of the wormhole geometry from Kerr \citep{urtubey25}. These results reinforce the conclusion that magnetic and rotational effects must be retained in the strong-field region, while also providing a physical basis for the magnetic-field terms used phenomenologically in the outer-flow discussion.

It should be noted that the magnetic field near a wormhole could theoretically be not only dipole but also monopole, without requiring the existence of a magnetic charge \citep{piotrovich24a}. This could also have a significant impact on the geometry of the jets. In particular, in the presence of an accretion disk, a monopole field could generate asymmetric or even one-sided jets \citep{piotrovich24a}.

\section{Discussion}

The analysis presented here is qualitative and phenomenological. A complete theoretical treatment would require: a self-consistent rotating wormhole metric, relativistic magnetohydrodynamic simulations, radiative transfer calculations, stability analysis of the wormhole throat, observational modeling of spectra and variability. The present fiducial estimates provide only scales; they do not constitute a numerical solution of the throat flow. Nevertheless, the proposed scenario demonstrates that wormhole-induced outflows could produce a broad range of astrophysical phenomena not expected in standard black hole accretion theory.

The present revision makes explicit that this conclusion does not follow from a Newtonian treatment of the throat. Rather, the Newtonian calculations provide a classification of the outer flow after the initial relativistic injection conditions have been established. The strong-field values of the specific energy, angular momentum, magnetic flux, and outflow velocity must be determined from a chosen rotating wormhole metric and a relativistic MHD calculation. In this sense, the present paper should be regarded as a phenomenological framework that identifies observable regimes and the physical quantities that a future metric-specific GRMHD calculation must predict.

The recent GRMHD simulations of Combi et al. provide an important consistency check: their simulations produce a hot rotating throat cloud and an outflow from the opposite side without the standard local accretion signature \citep{combi24}. Their result is qualitatively compatible with an unbound or weakly bound outward-flowing state, but it does not select the present regimes because the mapping depends on the metric and injected integrals of motion.

\section{Conclusion}

We have considered a hypothetical scenario in which matter accreted through one mouth of a Kerr-type wormhole emerges from the opposite mouth as an equatorial outflow. The revised classification is based on the specific energy and angular momentum of the emerging plasma. Unbound motion corresponds to $\epsilon>0$, while bound motion has $\epsilon<0$ and can either circularize through dissipation or return to the throat if circularization is inefficient. These structures differ from conventional accretion disks and may possess unusual observational properties.

The main quantitative scales obtained for a fiducial $10^8M_\odot$ system with $\dot M=0.1M_\odot\,\mathrm{yr}^{-1}$ are $L_{\rm kin}\sim10^{43}$--$10^{44}\,\mathrm{erg\,s^{-1}}$, $\rho\sim10^{-15}\,\mathrm{g\,cm^{-3}}$ at $20r_g$, and a representative flux-conserved field of order $10^2$ G at the same radius for the illustrative normalization adopted above. These numbers are scaling estimates, not predictions.

Thus, several potentially observable properties may distinguish wormhole-induced disk outflows from ordinary accretion disks, but none is unique in isolation:
\begin{itemize}
\item Dominance of radial over azimuthal velocities in the unbound component.
\item Temperature profiles controlled by expansion, shocks, and irradiation rather than by the standard thin-disk law.
\item Radiative recombination continua whose strength scales with the emission measure and ionization state.
\item Highly non-Keplerian turbulent structures during the circularization phase.
\item Transient contracting structures after the injected outflow decreases.
\item X-ray emitting fallback plasma shells.
\item Jets or Poynting-flux outflows in the absence of a locally obvious thin disk, with the important caveat that radiatively inefficient or obscured black hole flows can produce similar phenomena.
\item Strong magnetic activity at a nominally quiescent mouth, including Poynting-flux-dominated outflows or magnetic variability without a locally visible accretion disk.
\item Possible deviations of the inner flow from geodesic motion caused by an intrinsic dipolar magnetic field.
\end{itemize}

We stress that these signatures are predictions of the phenomenological scenario and are not unique wormhole diagnostics. Tidal-disruption debris can undergo eccentric self-intersection and circularization \citep{hayasaki13}, while radiatively inefficient black hole accretion flows can produce hot, optically thin, non-Keplerian structures and jets \citep{hada24,balakrishnan24}. Horizonlessness is likewise not unique to wormholes. A convincing identification would therefore require a combination of dynamical, spectral, magnetic, and spacetime-signature constraints. Their actual detectability must be established by future relativistic radiative-transfer and GRMHD calculations.

\authorcontributions{Conceptualization, M.P.; methodology, M.P.; software, M.P.; validation, S.B. and T.N.; formal analysis, M.P.; investigation, M.P., S.B. and T.N.; resources, M.P., S.B. and T.N.; data curation, S.B., T.N.; writing—original draft preparation, M.P.; writing—review and editing, M.P.; visualization, M.P.; supervision, M.P.; project administration, M.P.; All authors have read and agreed to the published version of the manuscript.}

\funding{This research received no external funding.}

\institutionalreview{Not applicable.}

\informedconsent{Not applicable.}

\dataavailability{The data underlying this article are available in the article.}

\acknowledgments{The authors express their gratitude to the reviewers for their very useful comments, which helped to significantly improve this paper.}

\conflictsofinterest{The authors declare no conflict of interest.}


\reftitle{References}

\externalbibliography{yes}
\bibliography{mybibfile}

\end{document}